\documentclass[12pt,A4paper]{article}
\makeatletter
\@addtoreset{equation}{section}
\makeatother

\usepackage[pdftex]{graphicx}
\usepackage{subfigure}
\usepackage{array}
\usepackage{amsmath}

\usepackage{rotating}
\usepackage{longtable}
\usepackage{mathtools}
\usepackage[numbers]{natbib}
\usepackage{array}
\usepackage{geometry}
\usepackage{color}
\usepackage{lipsum}

\usepackage{diagbox} %%%?????The
\usepackage{pdfpages}
\usepackage{multicol}
\usepackage[utf8]{inputenc}
\usepackage[english]{babel}
\usepackage{mathtools}
\usepackage{bbm}
\usepackage{amssymb}
\usepackage{amsthm}
\usepackage{hyperref}
\usepackage{bm}
\usepackage{mathtools}
\usepackage{fancybox}
\usepackage{multirow}
\usepackage{multicol}
\usepackage{caption}
\usepackage{subfigure}
\usepackage{algorithm}
\usepackage[affil-it]{authblk}
\usepackage[noend]{algpseudocode}
\usepackage{setspace}
\makeatletter
\newcommand*{\rom}[1]{\expandafter\@slowromancap\romannumeral #1@}
\newcommand{\tabincell}[2]{\begin{tabular}{@{}#1@{}}#2\end{tabular}}  
\makeatletter
\def\BState{\State\hskip-\ALG@thistlm}
\makeatother
\DeclarePairedDelimiter\abs{\lvert}{\rvert}%
\allowdisplaybreaks[4]

\usepackage{xcolor}
\renewcommand{\raggedright}{\leftskip=0pt \rightskip=0pt plus 0cm}
\title{{\bf Coefficient Shape Transfer Learning\\ for Functional Linear Regression}}
\author[1]{Shuhao Jiao\thanks{Corresponding Author: shuhao.jiao@cityu.edu.hk}}
\author[2]{Ian W.\ McKeague}
\affil[1]{Department of Biostatistics\\City University of Hong Kong}
\affil[2]{Department of Biostatistics\\ Columbia University}
\date{}
\begin{document}
	\maketitle			
	\setlength\parindent{0pt}
	\setlength{\parskip}{1em}
	\theoremstyle{definition}
	\newtheorem{theorem}{Theorem}
	\newtheorem{ass}{Assumption}
	\newtheorem{lemma}{Lemma}
	\newtheorem{remark}{Remark}
	\newtheorem{prop}{Proposition}
	\newtheorem{Definition}{Definition}
	\newtheorem{cor}{Corollary}
\begin{abstract}
The shapes of functions provide highly interpretable summaries of their trajectories. This article develops a novel transfer learning framework for functional linear regression that leverages coefficient shape information to overcome data scarcity. The proposed methodology integrates samples from the target model (target domain) with those from auxiliary models (source domains), enabling effective knowledge transfer across heterogeneous data sources. This shape-based transfer learning framework enhances robustness and generalizability: by being invariant to covariate scaling and signal strength, it ensures reliable knowledge transfer even when data from different sources differ in magnitude, and by formalizing the notion of coefficient shape homogeneity, it extends beyond traditional coefficient-equality assumptions to incorporate information from a broader range of source domains. We establish non-asymptotic convergence rates and minimax optimality of the estimator, showing that improvement depends on shape similarity, coefficient magnitude, and spectral decay rates of the covariance operators of covariates across different domains. A data-driven source domain selection procedure is developed to identify informative auxiliary domains. Simulations and a real data application to physical activity data from the U.S. National Health and Nutrition Examination Survey demonstrate the framework’s accuracy, robustness, and interpretability. \\

\noindent{\bf Keywords}: Coefficient shape, Functional linear regression, Minimax optimality, Non-asymptotic convergence, Transfer learning.
\end{abstract}	
\newpage
\section{Introduction}
\subsection{Background}
Machine learning techniques have achieved significant success in many practical problems. However, a common assumption in machine learning is that both training and test data come from the same distribution. Consequently, when new data from a different distribution is encountered, statistical models often need to be redeveloped from scratch. In many scenarios, re-collecting training data and rebuilding models is costly or impractical. Therefore, there is a growing demand to utilize existing datasets to aid in the development of new models, leading to the emergence of transfer learning. Transfer learning leverages knowledge gained from solving auxiliary problems and applies it to a new one. This technique draws inspiration from the human ability to apply previously acquired knowledge to improve performance on new tasks. Transfer learning has been applied in many fields, such as neuroimaging analysis (see e.g., \cite{ardalan2022transfer}), image classification (see e.g., \cite{ganin2015unsupervised,long2015learning}), natural language processing (see e.g., \cite{devlin2019bert}), protein function prediction (see e.g., \cite{heinzinger2019modeling}), functional data analysis (see e.g., \cite{cai2024transfer}, \cite{lin2024hypothesis}, \cite{hu2025transfer}, \cite{liu2025similarity}), and high-dimensional regression (see e.g., \cite{bastani2021predicting}, \cite{li2022transfer} and \cite{tian2023transfer})

Functional data analysis (FDA) encompasses a variety of statistical methods used to analyze data represented as functions, images, shapes, or more general objects (see e.g., \cite{wang2016functional}). 
Coefficient homogeneity has been employed to build parsimonious models (see, e.g., \cite{shen2010grouping}, \cite{ma2017concave}, and \cite{she2022supervised}), and can be readily adopted in transfer learning by assuming that models across different domains share similar coefficients. {Two pioneering papers, \cite{lin2024hypothesis} and \cite{hu2025transfer}, have made important contributions along this line of research.  
{\it A major and critical drawback} of the coefficient homogeneity assumption is its lack of robustness to data scaling: scaling can dramatically distort the differences between associated coefficients, leading to inconsistent and unreliable transfer of knowledge on coefficient amplitude across domains. This issue is particularly pronounced in transfer learning because source domains with high data variability can easily dominate those with lower variability, making data scaling unavoidable in many practical applications. Moreover, it reduces generalization and interpretability, as the equality constraint suppresses structural variations across domains—treating all coefficients as identical prevents the model from capturing shared but non-identical association patterns. 
In contrast, this paper develops a new transfer learning methodology based on {\it coefficient shape similarity}, which effectively overcomes the inherent weaknesses of the coefficient homogeneity assumption. It is worth noting that, while both \cite{lin2024hypothesis} and our work study the impact of signal strength on transfer learning, their work builds on coefficient homogeneity whereas ours emphasizes coefficient shape similarity, and the two works offer complementary insights on the transferred knowledge in the setting of functional linear model.}

\subsection{Problem of Interest}
We now present the problem setup. Since the intercept does not affect the convergence rate of the coefficient functions (see e.g., \cite{du2014penalized}), for simplicity we assume that the mean of response and functional covariate variables are both zero, and the target model of interest is given by
\begin{equation}
\label{target-model}
y_n^{(0)}=\langle X^{(0)}_{n},\beta^{(0)}\rangle+\epsilon^{(0)}_n,\ n=1,\ldots,N_0,\ \epsilon^{(0)}_n\overset{i.i.d.}\sim \mbox{subG}(\sigma^2),
\end{equation}
in which $\epsilon^{(0)}_n\sim \mbox{subG}(\sigma^2)$ indicates that $\epsilon^{(0)}_n$ follows a sub-Gaussian distribution with variance proxy $\sigma^2$, and are independent across $n$. Here $\langle \cdot,\cdot\rangle$ denotes the inner product in $L^2[0,1]$, $X_n^{(0)}(t)$ denotes the square-integrable functional covariate of the target model, and $\beta^{(0)}(t)$ denotes the associated unknown coefficient function. In addition, assume that we have data generated from $p$ auxiliary models as follows,
\begin{equation}
\label{auxiliary-model}
y_n^{(j)}=\langle X^{(j)}_{n},\beta^{(j)}\rangle+\epsilon^{(j)}_n,\ n=1,\ldots,N_j,\ \epsilon^{(j)}_n\overset{i.i.d.}\sim \mbox{subG}(\sigma^2),\ j=1,\ldots,p,
\end{equation}
where $\beta^{(j)}(t)$ is the unknown coefficient functions of the $j$'s auxiliary model. It is assumed that for some $j$'s, the function $\beta^{(j)}(t)$ shares a similar coefficient shape with $\beta^{(0)}(t)$, and the set of these domains, denoted by $\mathcal{A}$, is termed the {\it informative set}.

A transfer learning procedure is composed of two steps: pre-training and fine-tuning. Generally, for some known functional $f(\cdot)$, let $\beta^{(0)}(t)=f(\beta^{c},a_0)(t)$, where $\beta^{c}(t)$ represents the transferred knowledge that is to be estimated from the data in $\mathcal{A}$ and the target domain in the pre-training step, and $a_0$ accounts for the divergence between $\beta^{(0)}(t)$ and $\beta^{c}(t)$, being estimated in the fine-tuning step. After obtaining $\hat{\beta}^{c}(t)$ and $\hat{a}_0$ in the two aforementioned steps, the final estimator of $\beta^{(0)}(t)$ is given by $\hat{\beta}^{(0)}(t)=f(\hat{\beta}^{c},\hat{a}_0)(t)$.

\subsection{Our Contributions}
The fundamental contribution of this paper is the development of a novel transfer learning methodology that leverages coefficient shape to enhance the estimation of the target functional linear regression model. Coefficient shape learning targets the geometry of functional effect patterns—peaks, supports, and trends of coefficient functions—rather than only their pointwise magnitudes. In functional linear models, the shapes are highly interpretable summaries of how an entire covariate trajectory drives an outcome.  In real-world settings—such as neuroimaging, motion monitoring, or wearable-sensor systems—individuals often exhibit common activity patterns. What
distinguishes them is the shape of their trajectories—when peaks occur, how sharply they rise or fall, and how long plateaus persist. A shape-based modeling approach captures precisely these structural
characteristics. By focusing on geometry rather than magnitude, it adapts naturally to differences in signal strength. Consequently, this approach not only enhances the performance of the target model but also ensures that the underlying association patterns remain comparable and interpretable across domains. In summary, by modeling, aligning, and transferring function shapes, we gain three major payoffs: (i) stability—shape information is more robust to data scaling and magnitude variation; (ii) interpretability—localized peaks and sustained plateaus translate into clear mechanistic claims; and (iii) generalization—in contrast to coefficient-equality–based methods, our shape-based transfer learning framework flexibly incorporates information from more source domains, enhancing adaptability and predictive accuracy.

The principle is that, if the $j$-th auxiliary model is close to the target model, the coefficient shape of the two models should be similar, in the sense that $\beta^{(j)}(t)\approx \mbox{const.} \,\beta^{(0)}(t)$. 
Clearly, while two functions are of the same shape if they are equal, two functions of the same shape are not necessarily identical.  Therefore, this new principle enables the incorporation of information from a broader range of source domains, thereby enhancing model estimation. More importantly, coefficient shape homogeneity is {\it invariant} to covariate scaling, addressing the crucial limitations of coefficient homogeneity principle discussed in Section \ref{back}. The developed transfer learning procedure is able to yield significantly greater improvements than merely increasing the target domain’s sample size. Due to the infinite-dimensional nature of functional data, the prediction mean squared error of the ordinary least squares estimator decays at a rate slower than $O(N_0^{-1})$. Under suitable conditions on the source domains, this rate can be improved to  $O((N_{\mathcal{A}}+N_0)^{-1})$ or faster, where $N_{\mathcal{A}}=\sum_{j\in\mathcal{A}}N_j$. 
In principle, when the source domains exhibit similar coefficient shapes, slow spectral decay, and large coefficient magnitudes, the proposed transfer learning method yields greater improvements than simply increasing the sample size in the target domain. 
For the case where $\mathcal{A}$ is unknown, a data-driven approach is developed to identify it. To this end, we introduce a novel shape misalignment measure, which quantifies the discrepancy in coefficient shapes between the source domains and the target domain. Based on this measure, we define the informative set as the collection of source domains exhibiting sufficiently small shape misalignment. 
The identification is formulated as a regularized optimization problem with a concave penalty, and we show that this approach yields consistent identification of the true informative set under some regularity conditions.

The rest of the paper is organized as follows. In Section \ref{s2}, we develop the new transfer learning methodology based on coefficient shape homogeneity, and we also establish the convergence rates of the final estimator and study the minimax optimality. In Section \ref{s3}, we develop an identification method for the unknown informative set, and investigate the consistency property of the identification result. Finite-sample properties of the proposed methods are investigated through simulation studies in Section \ref{s4}. A case study of the association between age and physical activity levels using occupation time curves as covariates is provided in Section \ref{s5}. Section \ref{s6} concludes the paper. Technical proofs and algorithms are given in the appendix.
\section{Coefficient Shape Transfer Learning}
\label{s2}
\subsection{Preliminaries}
The coefficient and covariate functions are assumed to belong to $L^2[0,1]$, in which the inner product is defined as $\langle x,y \rangle=\int_0^1x(t)y(t)\,dt$, and the $L^2$-norm is defined by $\|x\|^2=\int_0^1x^2(t)\,dt<\infty$. Suppose that there is a target data source including $N_0$ observations $\{(X^{(0)}_n(t),y^{(0)}_n)\colon n=1,\ldots,N_0\}$ following the model \eqref{target-model}, and, in addition, there are $p$ auxiliary data sources, where in the $j$-th source the $N_j$ observations $\{(X^{(j)}_n(t),y^{(j)}_n)\colon n=1,\ldots,N_j\}$ follow model \eqref{auxiliary-model}.  The (empirical) covariance operator of the design covariates $\{X^{(j)}_n(t)\colon n= 1,\ldots, N_j\}$ is defined as $\mathcal{C}_j(x)(t)=\int C_j(t,s)x(s)ds$, 
where $C_j(t,s)=N^{-1}_j\sum_{n=1}^{N_j}\{X^{(j)}_n-\bar{X}^{(j)}\}(t)\{X^{(j)}_n-\bar{X}^{(j)}\}(s)$, and $\bar{X}^{(j)}(t)=N^{-1}_j\sum_{n=1}^{N_j}X_n^{(j)}(t)$, for $j=0,\ldots, p$. For simplicity, we assume that $\bar{X}^{(j)}(t)=0$ for all $j$'s. It is readily checked that each $\mathcal{C}_j(\cdot)$ is a compact operator on $L^2[0,1]$.

Our objective  is to fit the target model $y_n^{(0)}=\langle X_n^{(0)},\beta^{(0)}\rangle+\epsilon_n^{(0)}$, where $\beta^{(0)}(t)\in L^2[0,1]$. By an extension of Mercer's theorem to compact covariance operators due to \citet{baker1981compact}, and the corresponding Karhunen--Lo\'eve representation theorem, we have the following functional principal component decompositions 
\begin{equation*}
\mathcal{C}_0(\cdot)=\sum_{d\ge1}\theta_{0d}\langle\nu_d,\cdot\rangle\nu_d(t),\ X_n^{(0)}(t)=\sum_{d\ge1}\xi^{(0)}_{nd}\nu_d(t)=\sum_{d\ge1}\langle X_n^{(0)},\nu_d\rangle\nu_d(t),
\end{equation*} 
where $\{(\theta_{0d},\nu_d(t))\colon d\ge1\}$ are the eigenpairs of $\mathcal{C}_0(\cdot)$. 
Although technically $\mathcal{C}_0(\cdot)$ depends on the  sample size $N_0$, as we argue in Remark \ref{remark3} below it is reasonable  to assume that the target coefficient function belongs to an  ``enhanced" parameter space (see also e.g., \citet{cai2006prediction}, \citet{hall2007methodology} and \citet{jiao2023functional}):
$$\Theta_\beta=\left\{\beta(t)\in L^2[0,1]\colon \beta(t)=\sum_{d\ge1}b_{0d}\nu_d(t), \sum_{d\ge1}b_{0d}^2<\infty\right\}.$$
%where the eigenfunctions $\{\nu_d(t)\colon d\ge1\}$  do not depend on the sample size.
More generally, \citet{yuan2010reproducing} and \citet{cai2012minimax} examined the scenario in which the reproducing kernel associated with the coefficient function is not necessarily well aligned with the covariance kernel of the functional covariate, but we do not pursue that direction here.
\begin{remark}
\label{remark1}
It is assumed that the variance of $\epsilon^{(j)}_n$ is the same for all $j=0,1,\ldots,p$. This assumption is reasonable in the context of the article for two reasons: 1) the error variance of different sources can always be scaled to a common level, and 2) the coefficient shape is invariant to data scaling.
\end{remark}
In the sequel, we use ``$c$" to denote generic positive constants. Let $a_n\asymp b_n$ denote $c^{-1}b_n<a_n\le cb_n$ for some $c>1$, $a_n\lesssim b_n$ denote $a_n\le cb_n$ for some $c>0$, and $a_n\gtrsim b_n$ denote $a_n\ge cb_n$ for some $c>0$. 

\subsection{Coefficient Shape Misalignment and Informative Set}
For the auxiliary models, assume that the coefficients and covariates admit the following basis representations
$\beta^{(j)}(t)=\sum_{d\ge1}b_{jd}\nu_d(t),\ X^{(j)}_{n}(t)=\sum_{d\ge1}\xi^{(j)}_{nd}\nu_d(t).$ %Although it may be more efficient to represent the auxiliary models using another set of basis functions, we still  recommend representing the models with $\{\nu_d(t)\colon d\ge1\}$, because this set of basis functions efficiently captures the target model, emphasizing components of the auxiliary models that are most relevant to the target one.
To evaluate coefficient shape discrepancy, we define the {\it coefficient shape misalignment} between $\beta^{(j)}(t)$ and $\beta^{(0)}(t)$ as $\mathcal{M}_j=\{M^{(j)}_{dd'}, 1\le d<d'\},\ \mbox{where}\ M^{(j)}_{dd'}=b_{jd}b_{0d'}-b_{jd'}b_{0d}.$ Notably, when $\|\mathcal{M}_j\|^2=\sum_{d<d'}\{M_{dd'}^{(j)}\}^2=0$, $\beta^{(j)}(t)= c \beta^{(0)}(t)$. %and $\beta^{(j)}(t)$ and $\beta^{(0)}(t)$ are considered to share the same shape. 
It is important to recognize that $\|\mathcal{M}_j\|$ can be small simply due to $\|\beta^{(j)}\|$ being small. To eliminate the influence of coefficient magnitude and ensure robust identification of the informative set, we define $\mathcal{A}$ with a threshold $\tilde{\lambda}\ge0$ as $$\mathcal{A}=\{j\colon \|\mathcal{M}_j\|\le\tilde{\lambda}\|\beta^{(j)}\|\|\beta^{(0)}\|\}.$$
 
The informative set is {\it robust} to covariate scaling because $\|\mathcal{M}_j\|/(\|\beta^{(j)}\|\|\beta^{(0)}\|)$ remains unchanged if the covariate functions are scaled to $c_0X_{n}^{(j)}(t)$ and $c'_0X_{n}^{(0)}(t)$, while accordingly the associated coefficient functions are scaled to $c^{-1}_0\beta^{(j)}(t)$ and ${c'}^{-1}_0\beta^{(0)}(t)$ with arbitrary constants $c_0,c'_0\in\mathbb{R}\setminus\{0\}$. The tuning parameter $\tilde{\lambda}$ controls the coefficient shape similarity between the target model and the auxiliary models, with a larger $\tilde{\lambda}$ leading to a more extensive informative set. It will be seen later that when $\tilde{\lambda}$ is sufficiently small, the information in $\mathcal{A}$ contributes to improving the estimation of $\beta^{(0)}(t)$. 
This parameter will be selected using a cross-validation method, as discussed in Section \ref{s3}.

\begin{remark}
Generally, the covariates and coefficients in the auxiliary models admit the orthogonal decompositions
$\beta^{(j)}(t)=\sum_{d\ge1}b_{jd}\nu_d(t)+\eta_j(t),\ X_n^{(j)}(t)=\sum_{d\ge1}\xi_{nd}^{(j)}\nu_d(t)+Z_n^{(j)}(t),$
where $\eta_j(t),Z_n^{(j)}(t)\in\Theta^\perp_\beta\subset L^2[0,1]$, and $\Theta^\perp_\beta$ is the orthogonal component of $\Theta_\beta$. The components $Z_n^{(j)},\eta_j$, assumed to be irrelevant to the target model, and are not incorporated in knowledge transfer. Thus, in practice, we use $\{\nu_d(t)\colon d\ge1\}$ as basis functions to construct approximations for $\beta^{(j)}(t)$ and $X_n^{(j)}(t)$, making the assumptions $\beta^{(j)}(t)=\sum_{d\ge1}b_{jd}\nu_d(t),\ X^{(j)}_{n}(t)=\sum_{d\ge1}\xi^{(j)}_{nd}\nu_d(t)$ reasonable.
\end{remark}
\subsection{Estimation with Transferred Coefficient Shape}
\label{est}
In this section, we develop the estimation procedure. We assume that the coefficient function of the target model admits the decomposition $\beta^{(0)}(t)=a_0\beta^c(t)$. Here $\beta^c(t)$ is termed the coefficient shape component, and $a_{0}$ is termed the coefficient amplitude component.  In addition, each coefficient function of the auxiliary models admits the decomposition 
$$\beta^{(j)}(t)=a_{j}\{\beta^{c}(t)+h_{j}(t)\},\ \mbox{where}\ \langle\beta^{c},h_{j}\rangle=0,\ h_{j}(t)\in\Theta_\beta.$$ 
This decomposition is not identifiable, because $\beta^{(j)}(t)=(ra_{j})[r^{-1}\{\beta^{c}(t)+h_{j}(t)\}]$ for arbitrary $r\ne0$. For identifiability and consistency, it is required that $\beta^{c}(t)=\sum\limits_{j\in\mathcal{A}\cup\{0\}}\omega_{j}\beta^{(j)}(t)$ if $h_{j}(t)=0$ for $j\in\mathcal{A}$ ({note that this does {\it not} mean we assume $h_{j}(t)=0$ for $j\in\mathcal{A}$}), and impose the restriction $\sum_{j\in\mathcal{A}\cup\{0\}}\omega_{j}a_{j}=1$. Under this restriction, we equivalently have that, if $h_{j}(t)=0$ for $j\in\mathcal{A}$, 
\begin{equation*}
\beta^{c}(t)=\sum_{j\in\mathcal{A}\cup\{0\}}\omega_{j}\sum_{d\ge1}b_{jd}\nu_d(t)\overset{\vartriangle}=\sum_{d\ge1}b_d^c\nu_d(t),
\end{equation*}
where $b_d^c=\sum_{j\in\mathcal{A}\cup\{0\}}\omega_{j}b_{jd}$. Note that the coefficient shape component $\beta^c(t)$ is shared by the target model and the auxiliary models, making it possible to use the samples in the source domains to improve the estimation of $\beta^c(t)$, provided that $\|a_jh_j\|$ are sufficiently small for $j\in\mathcal{A}$. Then we adjust the amplitude of the target coefficient to get the final estimator of $\beta^{(0)}(t)$. Now we present the transfer learning procedure: 

{\bf Pre-training step}. We estimate the coefficient functions of the target and related auxiliary models
\begin{equation*}
\check{\beta}^{(j)}(t)=\arg\min\limits_{\beta(t)\in\Theta_\beta}\sum\limits_{n=1}^{N_{j}}\left\{y^{(j)}_{n}-\langle X^{(j)}_{n},\beta\rangle\right\}^2,\ j\in\mathcal{A}\cup\{0\}.
\end{equation*}
The least squares estimation is performed after projection onto the basis $\{\nu_d(t) \colon d \ge 1\}$ (see e.g., \citet{cai2006prediction} and \citet{hall2007methodology}). 
Using the basis representation, we derive the multivariate regression form of the functional model $y^{(j)}_n=\sum_{d\ge1}\xi^{(j)}_{nd}b_{jd}+\epsilon_n^{(j)}=\sum_{d=1}^D\xi^{(j)}_{nd}b_{jd}+\tilde{\epsilon}_n^{(j)}$, where $\tilde{\epsilon}_n^{(j)}=\epsilon_n^{(j)}+\sum_{d\ge D+1}\xi^{(j)}_{nd}b_{jd}$ for some sufficiently large $D\ge1$. Then we use the least squares method to obtain the pre-estimate $\check{\bm{b}}_j=(\check{b}_{j1},\ldots,\check{b}_{jD})$, and the resulting estimate of $\beta^{(j)}(t)$ is given by $\check{\beta}^{(j)}(t)=\sum_{d=1}^D\check{b}_{jd}\nu_d(t)$.
Then we estimate the coefficient shape component as follows 
$$\hat{\beta}^{c}(t)=\sum_{j\in\mathcal{A}\cup\{0\}}\omega_j\check{\beta}^{(j)}(t),\ \omega_{j}=\mbox{sign}(\langle\check{\beta}^{(j)},\check{\beta}^{(0)}\rangle)\times\frac{N_{j}}{N_\mathcal{A}+N_0},$$
where $N_\mathcal{A}=\sum_{j\in\mathcal{A}}N_j$.  We estimate the coefficient function separately in each domain, eliminating the need to re-estimate $\beta^c(t)$ from scratch when incorporating a new source domain, thereby significantly reducing the computational burden.

{\bf Fine-tuning step}. We estimate the coefficient amplitude based on $\hat{\beta}^c(t)$ as follows 
\begin{equation*}
\hat{a}_0=\arg\min\limits_{a_0\in\mathbb{R}}\sum\limits_{n=1}^{N_{0}}\left\{y^{(0)}_{n}-a_0\langle X^{(0)}_{n},\hat{\beta}^{c}\rangle\right\}^2.
\end{equation*}
Then %we %adjust the pre-estimate of the target model to obtain 
the final estimator of the target coefficient function is $\hat{\beta}^{(0)}(t)=\hat{a}_0\hat{\beta}^{c}(t)$.

In the pre-training step, we estimate the coefficient shape component using all samples from the source domains within the informative set, and $\hat{\beta}^c(t)$ serves as a reliable estimator of $\beta^c(t)$ when $\tilde{\lambda}$ is small. The probability limit of $\hat{\beta}^{c}(t)$ is $\sum_{j\in\mathcal{A}\cup\{0\}}\omega_j\beta^{(j)}(t) = \beta^c(t) + \sum_{j\in\mathcal{A}\cup\{0\}}\omega_j a_j h_{j}(t).$ As a special case, when $\tilde{\lambda} = 0$, the estimator $\hat{\beta}^c(t)$ is consistent for $\beta^c(t)$ as $N, D \to \infty$. Notably, when the source domains are sufficiently informative and $N_{\mathcal{A}} \gg N_0$, the convergence rate of $\hat{\beta}^c(t)$ is expected to be substantially faster than that of the ordinary least squares estimator using only target domain samples. In the fine-tuning step, we further estimate the coefficient amplitude. The estimator $\hat{a}_0$ achieves a convergence rate of $O_p(N_0^{-1/2})$, which again surpasses that of the ordinary least squares estimator on the target domain alone, owing to the fact that $a_0$ is a scalar. Therefore, with a properly chosen $\tilde{\lambda}$, the convergence rate $\hat{\beta}^{(0)}(t)$ is expected to be significantly faster the estimators based solely on the target domain. A detailed discussion is provided below.

\subsection{Theoretical Properties of the Final Estimator}
\label{convergence}

\subsubsection{Homogeneous Design}
\label{homo}

For simplicity in deriving the convergence rate of the transfer learning estimator, in this section we assume that there is only one source domain in $\mathcal{A}$ (say, $\mathcal{A}=\{1\}$), and $\mathcal{C}_1(\cdot)$ and $\mathcal{C}_0(\cdot)$ are perfectly aligned, that is, $\mathcal{C}_1(\cdot)=\sum_{d\ge1}\theta_{1d}\langle\nu_d,\cdot\rangle\nu_d(t)$ with $\theta_{11}\ge\theta_{12}\ge\cdots$ being the eigenvalues of $\mathcal{C}_1(\cdot)$. Notationally,
let $\|\hat{\beta}^{(0)}-\beta^{(0)}\|^2_{\mathcal{C}_0}=\langle\hat{\beta}^{(0)}-\beta^{(0)},\mathcal{C}_0(\hat{\beta}^{(0)}-\beta^{(0)})\rangle$ to evaluate the prediction error. Here, $\|\beta^{(0)}\|$ is treated as fixed, whereas the magnitude of the source domain coefficient $\|\beta^{(1)}\|$ is allowed to vary.

To better understand the influence of source domain on the overall convergence rate, we consider the regime in which the source domain plays a dominant role in determining the convergence rate. To streamline the discussion, we assume that the design functions $\{X_n^{(j)}(t) \colon j \in \mathcal{A} \cup \{0\}\}$ are non-random to underscore the core factors governing convergence behavior. The extension to the random design setting is presented in the supplementary material.

We first make the following assumptions.
\begin{itemize}
\item[(A1)] For $j=0,1$, $\{\epsilon^{(j)}_n\colon n=1,\ldots,N_j\}$ are independent zero-mean and sub-Gaussian with variance proxy $\sigma^2$, meaning that $\mbox{P}\{\abs{\sum_{n=1}^{N_j}s_n\epsilon_n^{(j)}}>t\}\le2\exp(-t^2/2\sigma^2\|\bm{s}\|^2)$ for any $\bm{s}=(s_1,\ldots,s_{N_j})$. %${E}[\exp(c\epsilon_n^{(j)})]\le\exp(\sigma^2c^2/2)$, for arbitrary $c\in\mathbb{R}$.
\item[(A2)] For $n\ge1,\ j=0,1$, $\xi^{(j)}_{nd}\lesssim d^{-\alpha_j/2}$ and $\theta_{jd}\asymp d^{-\alpha_j}$ with $\alpha_j>1$, and $b_{0d}\lesssim d^{-\tilde{\alpha}_0}$ with $\tilde{\alpha}_0>1/2$. 
\end{itemize}
Assumption (A1) posits sub-Gaussian random noise in both the primary and informative auxiliary samples, along with a finite second moment for the response vector. 
Assumption (A2) specifies the decaying rate of the covariate scores and the coefficient template scores. The assumption that $\alpha_j > 1$ and $\tilde{\alpha}_j > 1/2$ arises from the requirement that all functional elements are square-integrable. 

\begin{remark}
\label{remark3}
As mentioned in the preliminary, it is reasonable to assume $\beta^{(0)}(t)\in\Theta_\beta$. It is mainly because the eigenfunctions $\{\nu_d(t)\colon d\ge1\}$ leads to stable scores as sample sizes diverge. 
%In assumption (A2), {the decay rates} of the coefficient scores and eigenvalues remain constant as the sample size increases. 
Under a random design and some mild conditions on $\alpha_j$, $\tilde{\alpha}_j$, it can be shown that $|{\theta}_{jd}-\tilde{\theta}_{jd}|\le\frac{1}{2}\tilde{\theta}_{jd}$  (this property is often used in existing literature, see e.g., \citet{cai2006prediction} and \citet{hall2007methodology}) and $|{b}_{jd}-\tilde{b}_{jd}|\le\frac{1}{2}\tilde{b}_{jd}$ asymptotically almost surely for $j=1,\ldots,D$, where $\tilde{\theta}_{jd}$ and $\tilde{b}_{jd}$ denote the counterparts of ${\theta}_{jd}$ and ${b}_{jd}$ constructed using the eigenfunctions of $\mbox{E}\{\mathcal{C}_0(\cdot)\}$. This implies that, as the sample size grows, the decay rates of these scores remain stable. In this section, we focus on the scenario where this stability holds, in order to highlight the primary factors influencing the convergence rate. 
It is found that, under some regularity conditions (e.g., when the coefficient functions are sufficiently smooth), the convergence rate obtained under the random design setting coincides with that under the fixed design. More details are provided in the supplementary material.%In such cases, the estimation error of the eigenpairs of $\mathcal{C}_0(\cdot)$ does not dominate the overall convergence rate.
\end{remark}

%\begin{theorem}
%\label{thm0}
%Under Assumption (A1) and (A2), for $r_1=2\sigma^2\log(2\delta^{-1})$, $r_2=8\sigma^2\{\log(6)+\log(\delta^{-1})\}$, $\delta>0$ and $\tilde{\lambda}\in[0,1)$, with probability greater than $1-2\delta$, it holds that
%\begin{align*}
%\|\hat{\beta}^{(0)}-\beta^{(0)}\|^2_{\mathcal{C}_0}
%&\lesssim\frac{a_0^2r_1\sum\limits_{j=0,1}|\omega_j|s_{\alpha_0-\alpha_j}(D)}{N_\mathcal{A}+N_0}+\tilde{\lambda}^2+\frac{r_2}{N_0}+D^{-(\alpha_0+2\tilde{\alpha}_0-1)},
%\end{align*}
%where 
%\begin{align*}
%s_\alpha(D)=\left\{
%\begin{array}{ccc}
%D^{\alpha+1},& &\mbox{if }\alpha>-1.\\
%\log(D),& &\mbox{if }\alpha=-1.\\
%D^{-1},& &\mbox{if }\alpha<-1.
%\end{array}
%\right.
%\end{align*}
%\end{theorem}

%Theorem \ref{thm1} establishes the convergence rate of  $\hat{\beta}^{(0)}(t)$. 
%The non-asymptotic upper bound provided in Theorem \ref{thm3} consists of four components:

%The key insight from Theorem \ref{thm3} is that when $\mathcal{C}_1(\cdot)$ exhibits a slow spectral decay rate and the sample size in the source domain is large enough (%Furthermore, it satisfies the condition for minimax optimality specified in Theorem \ref{thm3}. 

%Regarding the minimax optimal property, we have the following result. 
We find that the transfer learning performance is affected by five factors: 1) the spectral decay rates of $\mathcal{C}_0(\cdot)$ and $\mathcal{C}_1(\cdot)$; 2) the coefficient magnitude in both target and source domain; 3) the coefficient shape divergence between the target and source domain; 4) the truncation error $\sum_{d\ge D+1}\xi^{(0)}_{nd}b_{0d}$; and 5) the sample sizes $N_\mathcal{A}$ and $N_0$. The following theorems encapsulate these findings on the convergence behavior of the final estimator, in which the function

\begin{align*}
g(x)=\left\{
\begin{array}{ccc}
%x^{\frac{\alpha_0+2\tilde{\alpha}_0-1}{\alpha_1-\alpha_0+1+(\alpha_0+2\tilde{\alpha}_0-1)}}=
x^{\frac{\alpha_0+2\tilde{\alpha}_0-1}{\alpha_1+2\tilde{\alpha}_0}},& &\mbox{if }\alpha_1-\alpha_0>-1\\
x\log D,& &\mbox{if }\alpha_1-\alpha_0=-1\\
x,& &\mbox{if }\alpha_1-\alpha_0<-1
\end{array}
\right.
\end{align*}
is used to recast the key factors, and $D$ is chosen to achieve the optimal convergence rate.

\begin{theorem}
\label{main-thm1}
Under Assumption (A1) and (A2), if $s_{\alpha_0-\alpha_1}(D)D^{\alpha_0+2\tilde{\alpha}_0-1}\asymp a_0^{-2}(N_\mathcal{A}+N_0)$ and $|\omega_0|D\le |\omega_1|s_{\alpha_0-\alpha_1}(D)$,  
then for any $\delta>0$, with probability greater than $1-3\delta$, it holds that
%Under Assumptions (A1) and (A2), if $|\omega_0|D\le |\omega_1|s_{\alpha_0-\alpha_1}(D)$, then the upper bound in Theorem 1 attains its minimum, given by
$$\|\hat{\beta}^{(0)}-\beta^{(0)}\|^2_{\mathcal{C}_0}\lesssim r_1g\left(\frac{a_0^2}{N_\mathcal{A}+N_0}\right)+\tilde{\lambda}^2\|\beta^{(0)}\|^2+\frac{r_1}{N_0},$$
where $r_1=2\sigma^2\log(2\delta^{-1})$.
\end{theorem}

Now we explain the imposed conditions. 
When $|\omega_0|D\le |\omega_1|s_{\alpha_0-\alpha_1}(D)$, the upper bound is primarily determined by the source domain, and when $s_{\alpha_0-\alpha_j}(D)D^{\alpha_0+2\tilde{\alpha}_0-1}\asymp a_0^{-2}(N_\mathcal{A}+N_0)$, the convergence rate of $\|\hat{\beta}^{(0)}-\beta^{(0)}\|^2_{\mathcal{C}_0}$ achieves the optimal convergence rate by balancing the bias (truncation error) and variance (estimation error).

Since $\|\beta^{(0)}\|$ is fixed, the result in Theorem \ref{main-thm1} yields the convergence rate $O(g(a_0^{2}(N_\mathcal{A} + N_0)^{-1}) \vee N_0^{-1} \vee \tilde{\lambda}^2)$. 
The function $g(\cdot)$ reflects how the difference in the coefficient magnitude and the spectral decay rates between the two domains influence the effectiveness of knowledge transfer. In particular, when the spectral decay rate of $\mathcal{C}_0(\cdot)$ is faster than that of $\mathcal{C}_1(\cdot)$, the benefit of transfer learning becomes more pronounced. For example, if $\alpha_1 - \alpha_0 < -1$ and $a_0$ is fixed, the convergence rate can be improved to $O((N_\mathcal{A} + N_0)^{-1})$. Conversely, if $\mathcal{C}_1(\cdot)$ exhibits a faster spectral decay rate than $\mathcal{C}_0(\cdot)$, the estimation accuracy may deteriorate due to a low signal-to-noise ratio along the leading functional principal components. 

%For example, the prediction error $\|\hat{\beta}^{(0)}-\beta^{(0)}\|^2_{\mathcal{C}_0}$ is decomposed into four components $$S_1=\frac{a_0^2\sum\limits_{j=0,1}|\omega_j|s_{\alpha_0-\alpha_j}(D)}{N_\mathcal{A}+N_0},\ S_2=\tilde{\lambda}^2,\ S_3=\frac{1}{N_0},\ S_4=\sum_{d\ge D+1}\xi^{(0)}_{nd}b_{0d}.$$The first component,  $S_1$ reflects the estimation error of the coefficient shape component  $\beta^c(t)$. The second component,  $S_2$ pertains to the bias caused by shape divergence between the source and target domains. It is noted that the bias is not only controlled by $\tilde{\lambda}$, which limits the maximum allowable shape discrepancy, but also by the target coefficient magnitude. The third component $S_3$ corresponds to the estimation error of the coefficient amplitude  $a_0$. The final component $S_4$ accounts for the truncation error of the target model. 

%The same discussion can be applied to the estimation error, and 
The following theorem establishes the convergence rate of the estimation error $\|\hat{\beta}^{(0)}-\beta^{(0)}\|^2$, and similar phenomena are also observed. 
\begin{theorem}
\label{main-thm2}
Under Assumption (A1) and (A2), let $D\asymp \{a_0^{-2}(N_\mathcal{A}+N_0)\}^{\frac{1}{\alpha_1+2\tilde{\alpha}_0}}$, if 
$$|\omega_0|D^{\alpha_0+1}\le |\omega_1|D^{\alpha_1+1}\ \mbox{and}\ a_0^2s_{\alpha_0-\alpha_1}(D)(N_\mathcal{A}+N_0)^{-1}+\tilde{\lambda}^2\ll1,$$  
then for any $\delta>0$, with probability greater than $1-3\delta$, it holds that
%Under Assumptions (A1) and (A2), if $|\omega_0|D\le |\omega_1|s_{\alpha_0-\alpha_1}(D)$, then the upper bound in Theorem 1 attains its minimum, given by
$$\|\hat{\beta}^{(0)}-\beta^{(0)}\|^2\lesssim r_1\left(\frac{a_0^{2}}{N_\mathcal{A}+N_0}\right)^{\frac{2\tilde{\alpha}_0-1}{\alpha_1+2\tilde{\alpha}_0}}+\left(\tilde{\lambda}^2+\frac{r_1}{N_0}\right)\|\beta^{(0)}\|^2,$$
where $r_1=2\sigma^2\log(2\delta^{-1})$.
\end{theorem}

The result in Theorem \ref{main-thm2} yields the convergence rate $O(\{{a_0^{-2}}(N_\mathcal{A}+N_0)\}^{-\frac{2\tilde{\alpha}_0-1}{\alpha_1+2\tilde{\alpha}_0}} \vee \tilde{\lambda}^2\vee N_0^{-1})$. The condition $|\omega_0|D^{\alpha_0+1} \le |\omega_1|D^{\alpha_1+1}$ ensures that the source domain dominates the estimation error. The condition $a_0^2s_{\alpha_0-\alpha_1}(D)(N_\mathcal{A}+N_0)^{-1}+\tilde{\lambda}^2\ll1$ restricts $\tilde{\lambda}$ and $\| \beta^{(1)} \|^{-1}$ from becoming excessively large. %Similarly, a faster spectral decay rate in the source domain leads to more substantial improvements in estimation accuracy.

Another insight from these theorems is that, including auxiliary models with weak signal strength (low coefficient magnitude) may increase estimation uncertainty. Specifically, the inclusion of such models reduces $\|\beta^c\|$  %due to the expression \eqref{beta_c}, 
resulting in a large value of $a_0$. For example, when $\abs{\omega_1}\approx1$ and $\tilde{\lambda}\approx0$, $a_0\approx \|\beta^{(0)}\|/\|\beta^{(1)}\|$. When $\beta^{(1)}(t)$ diminishes, the corresponding source domain provides negligible information, and its inclusion may inflate estimation error rather than reduce it. The last term in the upper bounds, $O(N_0^{-1})$, serves as the baseline error. As $\|\beta^{(j)}\| \to \infty$ and $\tilde{\lambda} \to 0$, the first two terms vanish, yet the overall estimation/prediction error does not diminish to zero accordingly due to the error introduced by estimating $a_0$.

%The convergence rate of the final estimator demonstrates clear improvement when $\tilde{\lambda}$ is not overly large and the size of the informative set $N_\mathcal{A}$ is sufficiently large. 

In the following theorems, we establish the minimax-optimal convergence rate in the transfer learning setting. Notationally, let  $r_a=a^2_1/a^2_0$.%Notationally, let $\lceil x\rceil$ denote the smallest integer greater than or equal to $x$.
\begin{theorem}
\label{thm3}
Suppose that the assumptions in Theorem \ref{main-thm1} hold, and 
\begin{equation}
\label{c1}
|\omega_0|+r_a(1-|\omega_1|)\lesssim r_a^{\frac{2\tilde{\alpha}_0}{\alpha_1+2\tilde{\alpha}_0}}(N_0+N_\mathcal{A})^{-\frac{\alpha_1}{\alpha_1+2\tilde{\alpha}_0}},
\end{equation}
then if $\tilde{\lambda}^2\lesssim N_0^{-\frac{\alpha_0+2\tilde{\alpha}_0-1}{\alpha_0+2\tilde{\alpha}_0}},$
the following result holds,
\begin{align*}
\inf_{\hat{\beta}^{(0)}}\sup_{\beta^{(0)}\in \Theta_\beta} \mbox{P}\left\{\|\hat{\beta}^{(0)}-\beta^{(0)}\|^2_{\mathcal{C}_0} \gtrsim \{r_a(N_\mathcal{A}+N_0)\}^{-\frac{\alpha_0+2\tilde{\alpha}_0-1}{\alpha_1+2\tilde{\alpha}_0}}+\tilde{\lambda}^2+N_0^{-1}\right\}>\frac{1}{4}.
\end{align*}
\end{theorem}

\begin{theorem}
\label{main-thm4}
Suppose that the assumptions in Theorem \ref{main-thm2}, along with condition \eqref{c1}, hold, 
%\begin{equation*}
%\label{c2}
%|\omega_0|+r_a(1-|\omega_1|)\lesssim r_a^{\frac{2\tilde{\alpha}_0}{\alpha_1+2\tilde{\alpha}_0}}(N_0+N_\mathcal{A})^{-\frac{\alpha_1}{\alpha_1+2\tilde{\alpha}_0}},
%\end{equation*}
then if $\tilde{\lambda}^2\lesssim N_0^{-\frac{2\tilde{\alpha}_0-1}{\alpha_0+2\tilde{\alpha}_0}}$,
the following result holds,
\begin{align*}
\inf_{\hat{\beta}^{(0)}}\sup_{\beta^{(0)}\in \Theta_\beta} \mbox{P}\left\{\|\hat{\beta}^{(0)}-\beta^{(0)}\|^2\gtrsim \{r_a(N_\mathcal{A}+N_0)\}^{-\frac{2\tilde{\alpha}_0-1}{\alpha_1+2\tilde{\alpha}_0}}+\tilde{\lambda}^2+N_0^{-1}\right\}>\frac{1}{4}.
\end{align*}
\end{theorem}

To analyze the impact of source domain, we impose the condition \eqref{c1}, under which the source domain predominantly influences the estimation and prediction in the target domain. %The assumption $\{r_a(N_\mathcal{A}+N_0)\}^{-\frac{\alpha_0}{\alpha_1+1}}\gtrsim g(a_0^{2}(N_\mathcal{A}+N_0)^{-1})$ imposes a joint restriction on the coefficient magnitude and spectral decay rate. 
The condition on $\tilde{\lambda}^2$ restricts the maximum shape misalignment from being excessively large. %The condition $N_0^{-\frac{2\tilde{\alpha}_0-1}{\alpha_0+2\tilde{\alpha}_0}}\gtrsim a_0^2N_0^{-1}$ restricts $\|\beta^{(1)}\|$ from being too small. 
This condition is natural; otherwise, estimation should rely solely on the target domain, since incorporating the source domain — when it exhibits large bias — would degrade the overall estimation accuracy. 

%As $\abs{\omega_1}$ is sufficiently large, 
Theorem \ref{main-thm4} implies that the convergence rate of estimation error in Theorem \ref{main-thm2} is minimax optimal under mild conditions. Notably, when $N_\mathcal{A}\gg N_0$, we have $\beta^c(t) \approx \beta^{(1)}(t)$, which implies $r_a \approx a_0^{-2}$. While the convergence rate of the prediction error is not guaranteed to be minimax optimal, it is so when $\alpha_1 > \alpha_0 - 1$. When $\alpha_1 = \alpha_0 - 1$, the quantity ${r_a(N_\mathcal{A}+N_0)}^{-\frac{\alpha_0+2\tilde{\alpha}_0-1}{\alpha_1+2\tilde{\alpha}0}}$ simplifies to approximately $a_0^{2}(N_\mathcal{A}+N_0)^{-1}$. This differs from the rate of the proposed estimator only by a logarithmic factor (i.e., a “$\log D$” term). In comparison, the minimax convergence rate of prediction error based solely on the target domain is $N_0^{-\frac{\alpha_0+2\tilde{\alpha}_0-1}{\alpha_0+2\tilde{\alpha}_0}}$. Even in the non-optimal case ($\alpha_1 < \alpha_0 - 1$), the final estimator still outperforms the minimax-optimal estimator based solely on the target domain — even when the sample size in the target domain is increased to $N_0 + N_\mathcal{A}$, as long as $\alpha_1 < \alpha_0$ and $r_a$ either converges to zero or remains bounded. This highlights the advantage of borrowing information from “high-quality” source domains rather than collecting new target-domain data, which may be expensive or impractical. 

\subsubsection{Heterogeneous Design}
\label{hetero}
In this section, we investigate the convergence rate of the final estimate  in the general scenario where $\{\mathcal{C}_j(\cdot)\colon j\in\mathcal{A}\}$ are not necessarily well aligned with $\mathcal{C}_0(\cdot)$ and there are multiple source domains. Let $\Sigma_j=N_j^{-1}\bm{\Xi}_j^{\top}\bm{\Xi}_j$, where $\bm{\Xi}_j=(\bm{\xi}^{(j)}_{1},\ldots,\bm{\xi}^{(j)}_{N_j})^{\top}$ being non-random, $\bm{\xi}^{(j)}_n=(\xi^{(j)}_{n1},\ldots,\xi^{(j)}_{nD})^{\top}$ and $\xi^{(j)}_{nd}=\langle X^{(j)}_{n},\nu_d\rangle$. Here we use the matrix $\Sigma^{-1/2}_j\Sigma_0\Sigma^{-1/2}_j$ to evaluate the spectral misalignment between $\Sigma_0$ and $\Sigma_j$ and denote $\rho_{j,max}$ as the maximal eigenvalue of $\Sigma^{-1/2}_j\Sigma_0\Sigma^{-1/2}_j$, 
%Particularly, $\rho_{0,max}=1$. 
%Suppose that, for $j\in\mathcal{A}$, $b_{jd}\lesssim a_ja_0^{-1}d^{-\tilde{\alpha}_j}$ and 
and let $\mathcal{R}_{j}=\sum_{d\ge D+1}\xi^{(j)}_{nd}b_{jd}$ for $j\in\mathcal{A}$ and $\mathcal{R}_0=0$. 
The following theorem establishes the convergence rate of the final estimate under the heterogeneous design. 
%We introduce the following additional assumption.
%\begin{itemize}
%\item[(A3)] For $n,j\ge1$, $\xi^{(j)}_{nd}\lesssim d^{-\alpha_j/2}$, and $\theta_{jd}\asymp d^{-\alpha_j}$ with $\alpha_j>1$, and $b_{jd}\lesssim a_jd^{-\tilde{\alpha}_j}$ with $\tilde{\alpha}_j>1/2$. 
%\end{itemize}

\begin{theorem}
\label{thm1}
Suppose that Assumptions (A1) and (A2) hold. 
Then for arbitrary $\delta>0$ and threshold $\tilde{\lambda}\in[0,1)$, with probability greater than $1-(|\mathcal{A}|+2)\delta$,
%\begin{align*} 
it holds that
$\|\hat{\beta}^{(0)}-\beta^{(0)}\|^2_{\mathcal{C}_0}\lesssim \widetilde{S}_1+\widetilde{S}_2+\widetilde{S}_3+\widetilde{S}_4$,
%\end{align*}
where 
\begin{align*}
\widetilde{S}_1&=a_0^2\sum\limits_{j\in\mathcal{A}\cup\{0\}}\omega^2_j\rho_{j,\max}\left(\frac{r_2}{N_j}+\mathcal{R}^2_j\right),\ \widetilde{S}_2=\tilde{\lambda}^2\|\beta^{(0)}\|^2,\ \widetilde{S}_3=\frac{r_1}{N_0},\ \widetilde{S}_4=D^{-(\alpha_0+2\tilde{\alpha}_0-1)},
\end{align*}
where $r_2=8\sigma^2(D\log6+\log\delta^{-1})$.   
\end{theorem}
In Theorem \ref{thm1}, %the four terms $\widetilde{S}_1, \widetilde{S}_2, \widetilde{S}_3, \widetilde{S}_4$ have the same interpretation as in Theorem \ref{main-thm1}.
the term $\rho_{j,\max}$ reflects the divergence in spectral decay rate between the $j$-th source domain and the target domain, and an additional term $\mathcal{R}_j^2$ arises in $\widetilde{S}_1$ because, under the heterogeneous design, the model truncation error is not necessarily orthogonal to the leading covariate scores. The first component,  $\widetilde{S}_1$ reflects the estimation error of the coefficient shape component  $\beta^c(t)$. The second component,  $\widetilde{S}_2$ pertains to the bias caused by shape divergence between the source and target domains. %It is noted that the bias is not only controlled by $\tilde{\lambda}$, which limits the maximum allowable shape discrepancy, but also by the target coefficient magnitude. 
The third component $\widetilde{S}_3$ corresponds to the estimation error of the coefficient amplitude  $a_0$. The final component $\widetilde{S}_4$ accounts for the truncation error of the target model.

\section{Informative Set Identification}
\label{s3}
\subsection{Identification through Regularization}
When $\mathcal{A}$ is unknown, we develop a data-driven procedure to identify it. In principle, the shape misalignment between $\beta^{(0)}(t)$ and $\{\beta^{(j)}(t)\colon j\in\mathcal{A}\}$ should be small, and thus we propose to  minimize the following objective function over  $\bm{\beta}(t)\overset{\vartriangle}=\{\beta^{(j)}(t)\colon j=0,\ldots,p\}$ 
\begin{equation*}
\label{objective}
S(\bm{\beta}(t),\lambda)=\frac{1}{2}\sum\limits_{j=0}^p\sum\limits_{n=1}^{N_j}\left\{y^{(j)}_{n}-\langle X^{(j)}_{n},\beta^{(j)}\rangle\right\}^2+\sum\limits_{j=1}^pJ_\lambda(\|\mathcal{M}_j\|),
\end{equation*}
where $J_\lambda(\cdot)$ is a concave penalty function such as the smoothly clipped absolute deviation penalty $J_\lambda(x)=\int_0^{|x|}\lambda\min\{1,\frac{(\gamma\lambda-t)_+}{(\gamma-1)\lambda}\}dt$ (SCAD, see \citet{fan2001variable}) and minimax concave penalty $J_{\lambda}(x)=\min\{\lambda|x|-{x^2}/{2\gamma},\gamma\lambda^2/2\}$ (MCP, see \citet{zhang2010nearly}). The concavity property is important for guaranteeing the consistency of informative set identification (see also \citet{jiao2024coefficient}, \citet{ma2017concave} and \citet{shen2010grouping}). The penalty term incorporates all the coefficient shape misalignments between the source domains and the target domain and shrinks insignificant coefficient shape misalignments, aiding in the identification of the informative set. %The regularized estimates of $\bm{\beta}(t)$ is $\widehat{\bm{b}}(t,\lambda)=\arg\min_{\bm{\beta}(t)}S(\bm{\beta}(t),\lambda)$. 

Since $S(\bm{\beta}(t),\lambda)$ involves elements of infinite dimensions, we minimize the following truncated version over $\{b_{jd}\colon d=1,\ldots,D, j=0,\ldots,p\}$ to obtain the regularized estimates $\{\hat{b}_{jd}\colon d=1,\ldots,D, j=0,\ldots,p\}$,
\begin{equation}
\label{trun-obj}
S_D(\bm{\beta}(t),\lambda)=\frac{1}{2}\sum\limits_{j=0}^p\sum\limits_{n=1}^{N_j}\left(y^{(j)}_{n}-\sum\limits_{d=1}^Db_{jd}\xi^{(j)}_{nd}\right)^2+\sum\limits_{j=1}^pJ_\lambda(\|\bm{M}_{j}\|),
\end{equation}
where $\bm{M}_{j}=\{M^{(j)}_{dd'}, 1\le d<d'\le D\}$. Then $\{\widehat{\bm{M}}_{j}\colon j\ge1\}$ from $\{\hat{b}^{(\lambda)}_{jd}\colon 1\le d\le D, j\ge0\}$ are constructed from the estimated coefficients as $\widehat{M}^{(j)}_{dd'}=\hat{b}^{(\lambda)}_{jd}\hat{b}^{(\lambda)}_{0d'}-\hat{b}^{(\lambda)}_{jd'}\hat{b}^{(\lambda)}_{0d}.$ We then identify the informative set as 
$\widehat{\mathcal{A}}=\{j\colon\|\widehat{\bm{M}}_{j}\|\le\tilde{\lambda}\|\hat{\bm{b}}^{(\lambda)}_{j}\|\|\hat{\bm{b}}^{(\lambda)}_{0}\|\},$ where $\hat{\bm{b}}^{(\lambda)}_{j}=(\hat{b}^{(\lambda)}_{j1},\ldots,\hat{b}^{(\lambda)}_{jD})$. The optimization problem in \eqref{trun-obj} can be solved using a linearized ADMM algorithm (see the supplementary material).

A grid search over $(\lambda, \tilde{\lambda})$ yields a sequence of candidate informative sets for $\widehat{\mathcal{A}}$. Each candidate is evaluated via a bootstrap procedure: the target coefficient $\beta^{(0)}(t)$ is estimated using transfer learning, residuals $\hat{\epsilon}_n = y^{(0)}_n - \hat{y}^{(0)}_n$ are computed, and bootstrap responses $\tilde{y}^{(0)}_n = \hat{y}^{(0)}_n + \hat{\epsilon}^*_n$ are generated, where $\hat{\epsilon}^*_n$ is obtained by resampling $\hat{\epsilon}_n$ with replacement. The model is then refitted to obtain a bootstrap estimate, and the mean squared prediction error (MSE) is calculated. This process is repeated (e.g., 1000 times) to compute the average MSE for each candidate set. The final informative set $\widehat{\mathcal{A}}$ is selected as the one minimizing the average MSE. After identifying the informative set, a post-selection procedure can be applied to remove source domains with significantly slower spectral decay rates and small coefficient magnitudes.

\subsection{Identification Consistency}
In this section, we investigate the consistency of the identified informative set. We establish the conditions on the tuning parameters, coefficient shape misalignment, and the dimension $D$ required to accurately find $\mathcal{A}$. We demonstrate that, under these regularity conditions, there exists a local minimizer of the objective function (\ref{trun-obj}) such that the associated informative set matches the true one with high probability. Before presenting the main theoretical results, we first introduce the following assumption:
\begin{itemize}
\item[(A3)] $J_\lambda(t)$ is a non-decreasing and concave function for $t\in [0, \infty)$ and $J_\lambda(0) = 0$. There exists a constant $\gamma>0$ such that $J_\lambda(t)$ is constant for all $t\ge \gamma\lambda\ge0$. The gradient $J_\lambda'(t)$ exists and is continuous except for a finite number of $t$ and $\lim_{t\to0+}J_\lambda'(t) = \lambda$.
\end{itemize}
Assumption (A3) holds for many concave penalty functions such as  SCAD penalty and MCP. By the results in Section \ref{convergence}, the tuning parameter $\tilde{\lambda}$ should converge to zero to ensure consistent final estimates. Notably, as $\tilde{\lambda}\to0$, $\beta^{(j)}(t)\to a_j\beta^c(t)$ for $j\in\mathcal{A}$. We show that under some regularity conditions there is a local minimizer of \eqref{trun-obj} around $\bm{B}^0=(\bm{b}^0_0,\ldots,\bm{b}^0_p)$, where $\bm{b}_j^0=(b^{0}_{j1},\ldots,b^{0}_{jD})$ and
\begin{equation*}
b^{0}_{jd}=\left\{
\begin{array}{ccc}
\langle a_j\beta^c,\nu_d\rangle,&&\mbox{if}\ j\in\mathcal{A}\cup\{0\},\\
\langle \beta^{(j)},\nu_d\rangle,&&\mbox{if}\ j\notin\mathcal{A}\cup\{0\},
\end{array}
\right.
\end{equation*}
so that the associated identified set coincides with the true informative set with high probability as the sample sizes $\{N_j\colon j\in\mathcal{A}\cup\{0\}\}$ diverge. Clearly, $\bm{B}^0$ is constructed in accordance with the definition of the informative set $\mathcal{A}$, and $\bm{b}^0_j$ shares exactly the same coefficient shape as $\bm{b}^0_0$ for $j\in\mathcal{A}$.  

Let $\{\bm{M}^0_{j}\colon j\ge1\}$ represent the coefficient shape misalignment associated with $\bm{B}^0$, and $\tau_N=\sqrt{N^{-1}\log N}$. The notation $\mathcal{R}_j$ is introduced in Section \ref{hetero}. The following theorem is about the consistency of $\widehat{\mathcal{A}}$.
\begin{theorem}
\label{thm5}
Under Assumptions (A1), (A2) and (A4), if the following two conditions hold,
\vspace{-2em}
\begin{itemize}
\item[C1).] $\|\bm{M}^0_{j}\|-2(\tau_{N_0}\tau_{N_j}+\tau_{N_0}\|\bm{b}^0_j\|+\tau_{N_j}\|\bm{b}^0_0\|)>\max\{\gamma\lambda,\tilde{\lambda}(\|\bm{b}^0_j\|+\tau_{N_j})(\|\bm{b}^0_0\|+\tau_{N_j})\},$ for $j\notin\mathcal{A}$,
\item[C2).] $\lambda\tilde{\lambda}\|\bm{b}^0_0\|\left(\sum_{j\in\mathcal{A}}\|\bm{b}^0_j\|\right)\gg \max\limits_{j=0,\ldots,p}(\log N_j+\sqrt{N_j\log N_j}\{\mathcal{R}_j+\tilde{\lambda}\|{\beta}^{(j)}\|\}),$
\item[C3).] $\frac{\sum_{i\in\mathcal{A}}a_{i}}{\sum_{i\in\mathcal{A}\cup\{0\}}a_{i}}-1<c<0$ and $\sum_{j\in\mathcal{A}}\left(\frac{\sum_{i\in\mathcal{A}\cup\{0\},i\ne j}a_{i}}{\sum_{i\in\mathcal{A}\cup\{0\}}a_{i}}-1\right)<c<0$,
\end{itemize}
\vspace{-1em}
%\end{equation*}
then, when $\{N_j\colon j=0,\ldots,p\}$ are sufficiently large, there exists a local minimizer of (\ref{trun-obj}) around $\bm{B}^0$ satisfying that $$\mbox{P}(\widehat{\mathcal{A}}\ne\mathcal{A})\le2D\sum_{j=0}^pN_j^{-1},$$ where $\widehat{\mathcal{A}}$ is the identified informative set associated with the minimizer.
\end{theorem}
Theorem \ref{thm5} shows that if $i \in \mathcal{A}$, the probability of excluding the $i$-th source domain from $\widehat{\mathcal{A}}$ decreases at a rate of $D/N_i$. As a result, if $D/N_i \to 0$ as $N_i \to \infty$, the probability of correctly including the $i$-th source domain in $\widehat{\mathcal{A}}$ converges to 1. 

Conditions C1) and C2) depend not only on the tuning parameters but also on the coefficient magnitudes. C1) highlights the crucial role of the concavity of $J_\lambda(\cdot)$. As the parameter $\gamma$ increases, the concavity of the penalty function $J_\lambda(\cdot)$ becomes less influential, potentially leading to a violation of C1). Concavity plays a crucial role in distinguishing substantial shape misalignments from minor variations by reducing over-shrinkage, thereby improving the accuracy in identifying the true informative set. C2) implies that, for a shape discrepancy to be detectable, it must be sufficiently pronounced relative to the coefficient magnitude, estimation error, and truncation error. $\mathcal{R}_j$ is removed from C2) under the homogeneous setting. C3) imposes a regularization constraint on the magnitude of the coefficients to avoid extremely large or small value of $\|\beta^{(0)}\|$ relative to the overall scale. 

\section{Simulation Studies}
\label{s4}
\subsection{General Setting}
In this section, we study the finite-sample properties of the developed methodologies through numerical experiments. Here we focus on the estimation error, as the main factors discussed in Section \ref{convergence} influence the prediction error in a similar manner. We omit the intercept, since it is not of major interest. We consider the following three settings.

{\bf Setting 1}.\ We simulate data from the functional regression models $$y^{(j)}_n=\langle X^{(j)}_{n},\beta^{(j)}\rangle+\epsilon^{(j)}_n,$$  
where $\epsilon^{(j)}_n\stackrel{i.i.d.}{\sim}\mathcal{N}(0,s^2)$. The functional covariates and the coefficient functions are generated from the following basis expansion,
%\begin{equation*}
$\beta_{j}(t)=\sum_{d=1}^Db_{jd}\nu_d(t),\ X^{(j)}_{n}(t)=\sum_{d=1}^D\xi^{(j)}_{nd}\nu_d(t),\ \xi^{(j)}_{nd}\sim\mathcal{N}(0,d^{-1.2\times2})$, 
%\end{equation*}
where $\{\nu_d(t)\colon d\ge1\}$ are Fourier basis functions. 

\begin{figure}[ht]
\center
\includegraphics[scale=0.16]{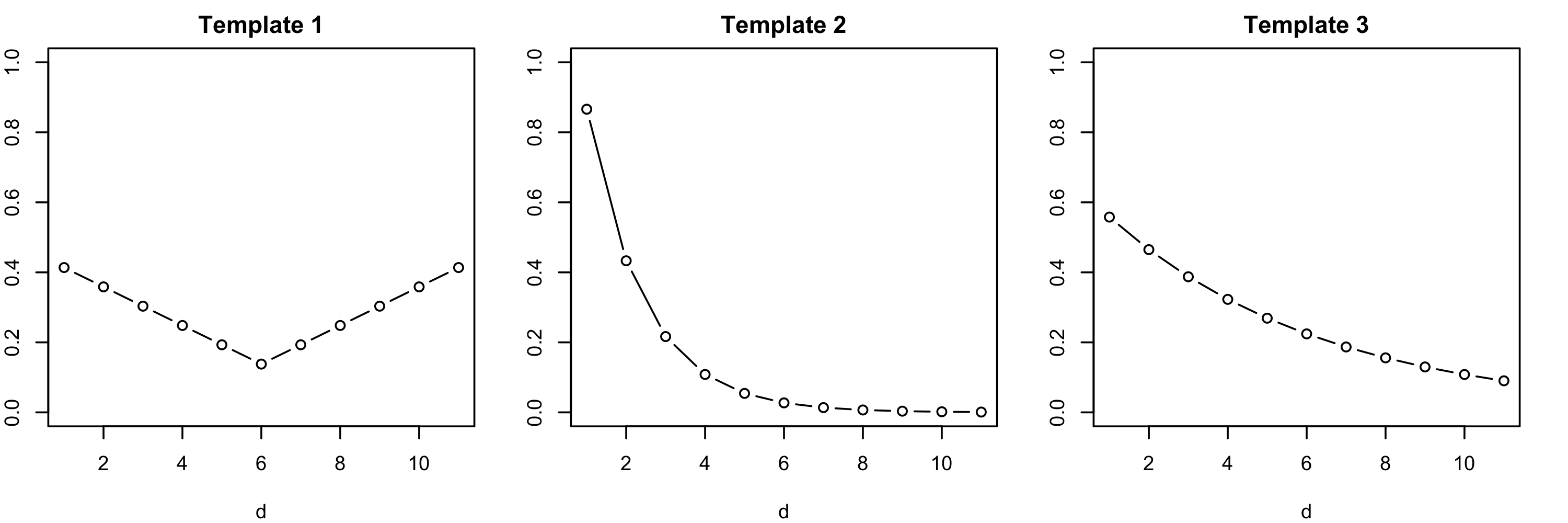}
\caption{Three templates coefficient scores in Setting 1 ($D=11$).} 
\label{cs}
\end{figure}

The coefficient scores $\{(b_{j1},\ldots,b_{jD}),\ j=0,\ldots,9\}$ are generated based on three different templates: $\mbox{Temp}_1=B_1/\|B_1\|$, $\mbox{Temp}_2=B_2/\|B_2\|$ and $\mbox{Temp}_3=B_3/\|B_3\|$, where
$B_1=\{3,3-\tau,\ldots,1+\tau,1,1+\tau,\ldots,3-\tau,3\},\ \tau=4/(D-1)$, $B_2=\{2^{-d}\colon d=1,\ldots,D\}$, and $B_3=\{1.2^{-d}\colon d=1,\ldots,D\}.$ The coefficient scores are then generated according to the following scheme:
\begin{itemize}
 \setlength\itemsep{-0.0em}
\item[For]$j=0,1,\ldots,5$: $\{b_{j1},\ldots,b_{jD}\}=f_j\times \mbox{Temp}_1$\hspace{3em} (V-shape)
\item[For]$j=6,7$: $\{b_{j1},\ldots,b_{jD}\}=f_j\times\mbox{Temp}_2$\hspace{3em} (fast-decay)
\item[For]$j=8,9$: $\{b_{j1},\ldots,b_{jD}\}=f_j\times\mbox{Temp}_3$\hspace{3em} (slow-decay)
\end{itemize}
where $\{f_j\colon j=0,1,\ldots,9\}=(3.0, 3.9, 4.8, 3.0, 3.9, 4.8, 3.6, 3.0, 3.9, 4.8)$. The three template scores for  $D = 11$  are shown in Figure \ref{cs}.

The coefficient function of the target model is determined by $B_1$. To investigate the impact of non-informative source domains, we designate the $6$-th to $9$-th source domains as non-informative. The oracle informative set consists of the $1$-st to $5$-th source domains.

{\bf Setting 2}.\ We simulate data from the following functional regression models 
\begin{align*}
y^{(0)}_n=\langle X^{(0)}_{n},\beta^{(0)}\rangle+\epsilon^{(0)}_n,\ n=1,\ldots,N, \\
y^{(1)}_n=\langle X^{(1)}_{n},\beta^{(1)}\rangle+\epsilon^{(1)}_n,\ n=1,\ldots,5N,
\end{align*}
where $\epsilon^{(j)}_n\stackrel{i.i.d.}{\sim}\mathcal{N}(0,s^2)$, $j=0,1$. The functional covariates $\{X^{(j)}_{n}(t)\colon j=0,1,\ n\ge1\}$ and the coefficient functions $\{\beta^{(j)}(t)\colon j=0,1\}$ are generated following the same basis expansion as in Setting 1, and $\xi^{(0)}_{nd}\sim\mathcal{N}(0,d^{-1.3\times2})$, $\xi^{(1)}_{nd}\sim\mathcal{N}(0,d^{-2\alpha})$, and $\{b_{j1},\ldots,b_{jd}\}=f_j\times\mbox{Temp}_4$, where $f_0=3.29,\ f_1=3.76$, and $\mbox{Temp}_4=B_4/\|B_4\|$, where $B_4=\{1.5^{-d}\colon d=1,\ldots,D\}$. 

{\bf Setting 3}.\ We simulate data from the following functional regression models 
\begin{align*}
y^{(0)}_n=\langle X^{(0)}_{n},\beta^{(0)}\rangle+\epsilon^{(0)}_n,\ n=1,\ldots,N, \\
y^{(1)}_n=\langle X^{(1)}_{n},\beta^{(1)}\rangle+\epsilon^{(1)}_n,\ n=1,\ldots,N,
\end{align*}
where $\epsilon^{(j)}_n\stackrel{i.i.d.}{\sim}\mathcal{N}(0,s^2)$, $j=0,1$. The functional covariates $\{X^{(j)}_{n}(t)\colon j=0,1,\ n\ge1\}$ and the coefficient functions $\{\beta^{(j)}(t)\colon j=0,1\}$ are generated following the same basis expansion as in Setting 1, and $\xi^{(j)}_{nd}\sim\mathcal{N}(0,d^{-1.2\times2})$ for $j=0,1$, and $\{b_{j1},\ldots,b_{jd}\}=f_j\times\mbox{Temp}_3$. We set $f_0=1$ and allow $f_1$ vary.

In Setting 1, we examine the impact of informative set identification on estimation performance. Specifically, we evaluate the transfer learning approach under two distinct scenarios: one where $\mathcal{A}$ is known and another where $\mathcal{A}$ is unknown. In Setting 2, we examine the impact of the decay rate of the covariate covariance operator. A smaller value of $\alpha$ is expected to result in a more significant improvement in estimation accuracy. In Setting 3, we study the influence of coefficient magnitude under varying magnitude of the coefficient in the source domain. In each setting, we repeat the simulation runs 1000 times to compute the average rooted mean squared error (RMSE), defined as $\{\int(\hat{\beta}^{(0)}-\beta^{(0)})^2\}^{1/2}$.

\subsection{Estimation Enhancement}
In this section, we focus on Setting 1 and consider three approaches and compare their estimation performance: the transfer learning approach based on coefficient shape homogeneity (TS), the transfer learning approach based on coefficient homogeneity (TE), and ordinary least squares estimation (OLS). To examine the impact of informative set identification, we implement the transfer learning approach using both the oracle informative set and the identified informative set. 

\begin{figure}[ht]
\center
\includegraphics[width=6.6in]{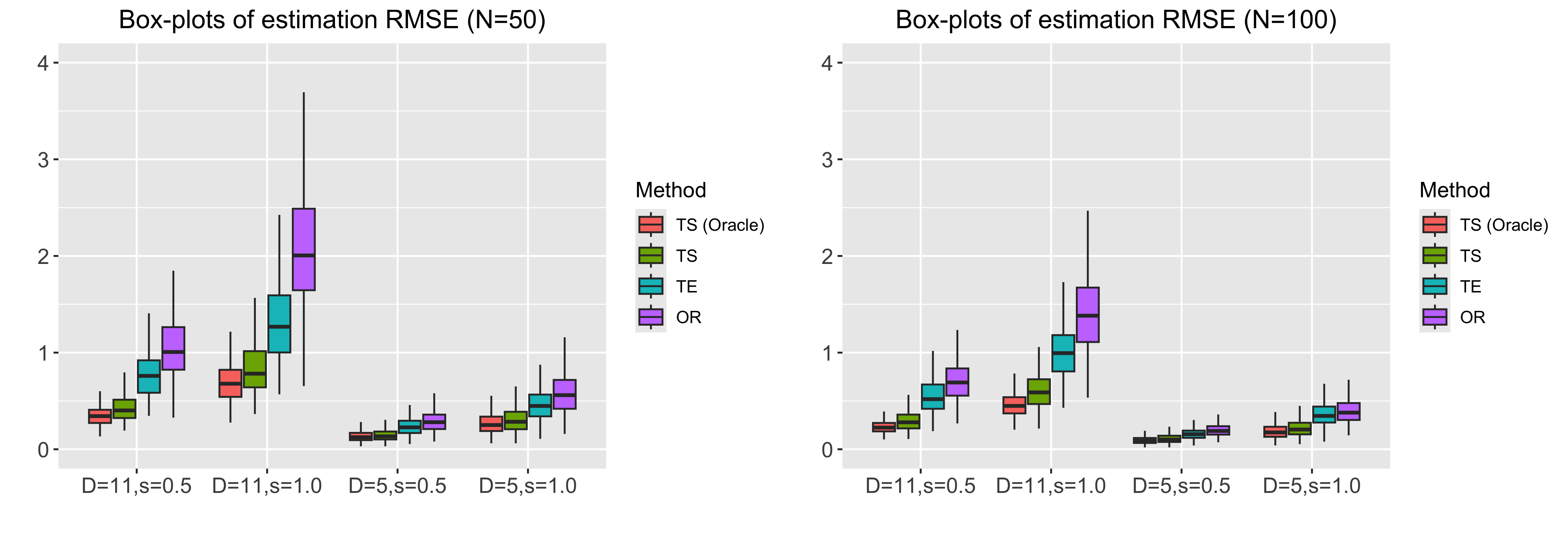}
\caption{Box-plots of estimation RMSE.} 
\label{box}
\end{figure}

To identify the informative set, we generate additional 200 validation samples and estimate the prediction error for each candidate informative set. The selected the set which yields the minimal RMSE, and the box-plots of the estimation RMSE are shown in Figure \ref{box}. In Table \ref{ci}, we present the average width of the estimation confidence bands at a significance level of 0.05, defined as $S_\alpha = \int \{u_\alpha(t) - l_\alpha(t)\} \mbox{dt}$, where $u_\alpha(t)$ and $l_\alpha(t)$ represent the upper and lower bounds, respectively, at the $0.05$ significance level.

\begin{table}[ht]
\centering
\captionsetup{justification=centering}
\caption{Width of 95\% confidence bands.}
\begin{tabular}{c|c|c|>{\centering\arraybackslash}p{0.8in}>{\centering\arraybackslash}p{0.8in}>{\centering\arraybackslash}p{0.8in}>{\centering\arraybackslash}p{0.8in}}
    
\hline 
    $N$  & $D$ & $s$ & TS (oracle) & TS &  TE & OLS\\
     \hline
    \multirow{4}{*}{50}&\multirow{2}{*}{5}					&0.5&0.547&0.585&1.024&1.209\\
         &&1.0&1.091&1.195&1.788&2.418\\
         \cline{2-7}
         &\multirow{2}{*}{11}&0.5&1.390&1.540&3.236&4.206\\
	 &&1.0&2.812&3.208&5.268&8.412\\
      \hline 
       \multirow{4}{*}{100}&\multirow{2}{*}{5}					&0.5&0.388&0.459&0.672&0.817\\
         &&1.0&0.777&0.929&1.441&1.635\\
         \cline{2-7}
         &\multirow{2}{*}{11}&0.5&0.922&1.159&2.308&2.863\\
	 &&1.0&1.851&2.161&4.065&5.726\\
      \hline 
\end{tabular}
\label{ci}
\end{table}

In this setting, the transfer learning estimator that utilizes the full oracle informative set achieves the best overall performance. This is expected, as the covariates across all source domains exhibit identical spectral decay rates and comparable coefficient magnitudes, creating an ideal environment for knowledge transfer. Notably, even without oracle knowledge, the transfer learning approach based on the identified informative set performs nearly as well, demonstrating only a slight degradation in performance due to the uncertainty involved in informative set selection.

Among all competing methods, the transfer learning estimator that leverages coefficient shape homogeneity achieves the best performance, as it flexibly captures shared structural patterns across source domains without requiring identical coefficient magnitudes. In contrast, the approach based on coefficient homogeneity imposes stricter assumptions, which limits its effectiveness when exact similarity does not hold. As expected, the ordinary least squares (OLS) estimator—serving as a baseline that ignores auxiliary information—consistently performs the worst, relying solely on limited target domain data.

These results highlight the advantages of incorporating shape-based transfer mechanisms and confirm the potential gains from accurate informative set identification in multi-source functional regression problems.

%In Setting 1, the transfer learning estimator based on the full oracle informative set achieves the best performance. This is because the covariates across all source domains share the same spectral decay rate, and the coefficient magnitudes are comparable. Even in this ideal scenario, the transfer learning approach using the identified informative set performs similarly to the oracle-based approach, with only slight inferiority due to the uncertainty in informative set identification. Among all methods, the transfer learning approach based on shape homogeneity performs the best, as it effectively leverages useful information from multiple sources. As expected, the ordinary least squares estimator, serving as the baseline, consistently performs the worst. %In the next section, we will demonstrate that including all source domains in the informative set is not always beneficial due to the variability in spectral decay rates.

\subsection{Spectral Decay Rate}
In this section, we examine the effect of the spectral decay rate of $\{\mathcal{C}_j(\cdot)\colon j\in\mathcal{A}\}$ and investigate Setting 2. We control the spectral decay rate of $\{\mathcal{C}_j(\cdot)\colon j\in\mathcal{A}\}$ by adapting the value of $\alpha$. A large value of $\alpha$ leads to a fast spectral decay rate. As discussed in Section \ref{homo}, an excessively fast spectral decay rate of covariates in the source domain may degrade estimation performance. The dimension $D$ is chosen as the smallest dimension, for which the cumulative proportion of the eigenvalues of $\mathcal{C}_0(\cdot)$ exceeds 90\%. 

The estimation RMSEs are displayed in Table \ref{sdr}, which justifies the theoretical findings in Section \ref{homo}. It is observed that when the spectral decay rate of the source domain is overly fast, incorporating its data may significantly degrade the estimation performance.  When $D = 11$, the deterioration in estimation accuracy due to the source domain is more pronounced, as more dimensions are truncated, leading to a significantly higher truncation bias. This occurs because the negligible variation in the tail covariate scores results in a high estimation error caused by the low signal-to-noise ratio. %These results highlight the importance of carefully selecting source domains based on the spectral decay rate of $\{\mathcal{C}_j(\cdot)\colon j\in\mathcal{A}\}$.

\begin{table}[ht]
\centering
\captionsetup{justification=centering}
\caption{Estimation RMSE under different spectral decay rates of $\mathcal{C}_1(\cdot)$. }
\begin{tabular}{c|c|c|>{\centering\arraybackslash}p{0.67in}>{\centering\arraybackslash}p{0.67in}>{\centering\arraybackslash}p{0.67in}>{\centering\arraybackslash}p{0.67in}>{\centering\arraybackslash}p{0.67in}|>{\centering\arraybackslash}p{0.67in}}    
\hline 
    \multirow{2}{*}{$N$}  &\multirow{2}{*}{$D$}  & \multirow{2}{*}{$s$}  & \multicolumn{5}{c|}{\tabincell{c}{TS}} &\multirow{2}{*}{OLS} \\
    \cline{4-8}
    & & &$\alpha=1.1$ &$\alpha=1.3$ & $\alpha=1.5$ &$\alpha=1.7$ &$\alpha=1.9$&\\
     \hline
    \multirow{4}{*}{50}&\multirow{2}{*}{5}&0.5&0.168&0.190&0.232&0.301&0.401&0.469\\
         &&1.0&0.337&0.380&0.464&0.599&0.794&0.938\\
         \cline{2-9}
         &\multirow{2}{*}{11}&0.5&0.400&0.531&1.040&2.221&4.425&1.545\\
	 &&1.0&0.748&1.048&2.049&4.318&7.233&3.081\\
      \hline 
       \multirow{4}{*}{100}&\multirow{2}{*}{5}&0.5&0.124&0.140&0.172&0.222&0.294&0.305\\
         &&1.0&0.247&0.281&0.344&0.443&0.586&0.609\\
         \cline{2-9}
         &\multirow{2}{*}{11}&0.5&0.267&0.364&0.731&1.599&3.313&0.980\\
	 &&1.0&0.489&0.731&1.447&3.119&6.000&1.951\\
      \hline 
\end{tabular}
\label{sdr}
\end{table}

\subsection{Coefficient Magnitude}
\begin{figure}[ht]
\center
\includegraphics[width=6in]{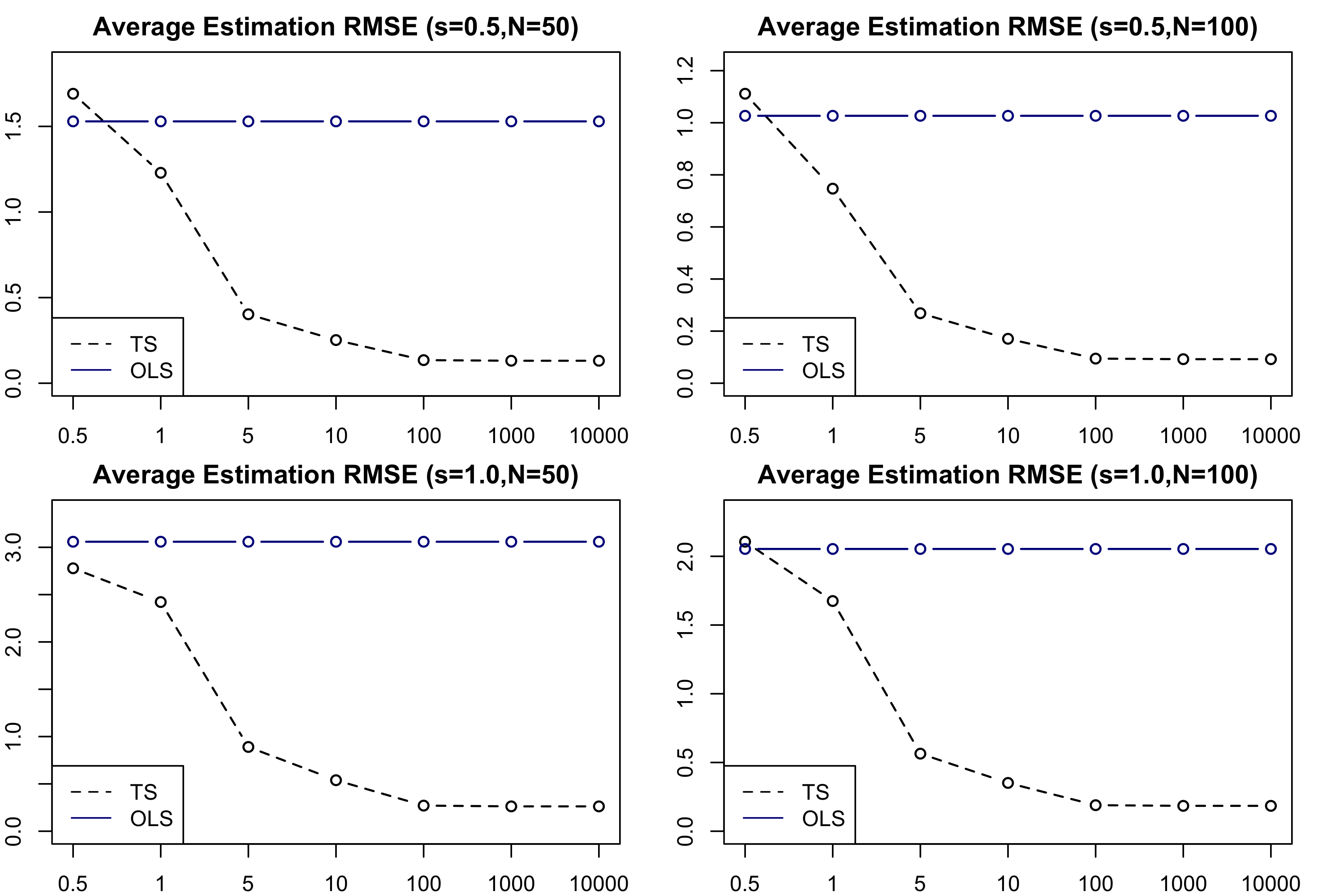}
\caption{Average Estimation RMSE ($D=11$). The x-axis represents the values of $f_1$.} 
\label{ampli}
\end{figure}

%As discussed in Section \ref{convergence}, if the coefficient magnitude in the source domains are small, the resulting scale coefficient will be high, learning to higher estimation error. 
In this section, we focus on Setting 3 to examine the performance of the transfer learning framework with varying $f_1$. The estimation RMSEs are displayed in Figure \ref{ampli}. The results suggest that a large value of $f_1$  leads to better estimation, primarily because the source domain helps reduce the scale coefficient. This phenomenon can be explained by the relationship between coefficient magnitude and signal strength. Specifically, a higher coefficient magnitude in the source domain enhances the signal-to-noise ratio, making it easier to extract meaningful information and improving the estimation accuracy in the target domain. Conversely, if the coefficient magnitude in the source domain is too small, the signal is then weak, limiting the benefits of transfer learning and potentially leading to suboptimal estimation in the target domain. As $f_1$ increases, the average RMSE does not vanish due to the estimation of $a_0$.

\section{Application}
\label{s5}
\subsection{Background on Wearable Device Data}
This paper is motivated by the increasing need to compare groups of subjects in terms of functional health outcomes, particularly in the context of physiological monitoring. Advances in wearable sensor technology have made it possible to collect vast amounts of longitudinal data over extended periods (ranging from weeks to months). These sensors are becoming indispensable in a wide array of research areas, including epidemiology, public health, and biomedical studies, such as those focused on congestive heart failure, pulmonary disease, diabetes, obesity, and Alzheimer’s disease.  Various functional data analysis methods have been developed to handle such data (see e.g., \citet{zhang2019review} and \citet{chang2022empirical}). 

Here we focus on data collected from wearable devices for monitoring physical activity, which has become a key area of interest in studies of human physiology and pathophysiology (see \citet{wright2017consumer}). Physical activity is often measured by the amount of time spent in various levels of intensity throughout the observation period, as assessed, for example, from data collected as ``counts" recorded by ActiGraph devices. Thresholds are commonly applied to sensor readings to define different activity categories (\citet{matthews2008amount}). However, these thresholds lack justification in the absence of separate validation studies. To address this concern, we move beyond measuring time within discrete activity categories and instead treat activity as a continuum. Specifically, we analyze activity in terms of occupation time, which represents the total duration spent above each activity level as a function of that level across the full range of sensor readings (see also \citet{chang2022empirical,chang2024concurrent}).  

We use scalar-on-function regression to investigate the association between chronological age and occupation time, using data from the 2005–2006 U.S.\ National Health and Nutrition Examination Survey (United States National Center for Health Statistics, 2005–2006). This study can help inform public health and medical research by highlighting how physical activity patterns vary across the lifespan, which could be useful for interventions or recommendations related to aging, health, and well-being. Here we split all subjects into 10 separate groups based on their gender and ethnic group. There are five ethnic groups in this dataset, including Mexican American, Other Hispanic, Non-hispanic White, Non-hispanic Black, and other ethnic groups (including multi racial), and two category of genders considered, including male and female.  

\subsection{Data Pre-processing}
\begin{figure}[ht]
\center
\includegraphics[width=6.2in]{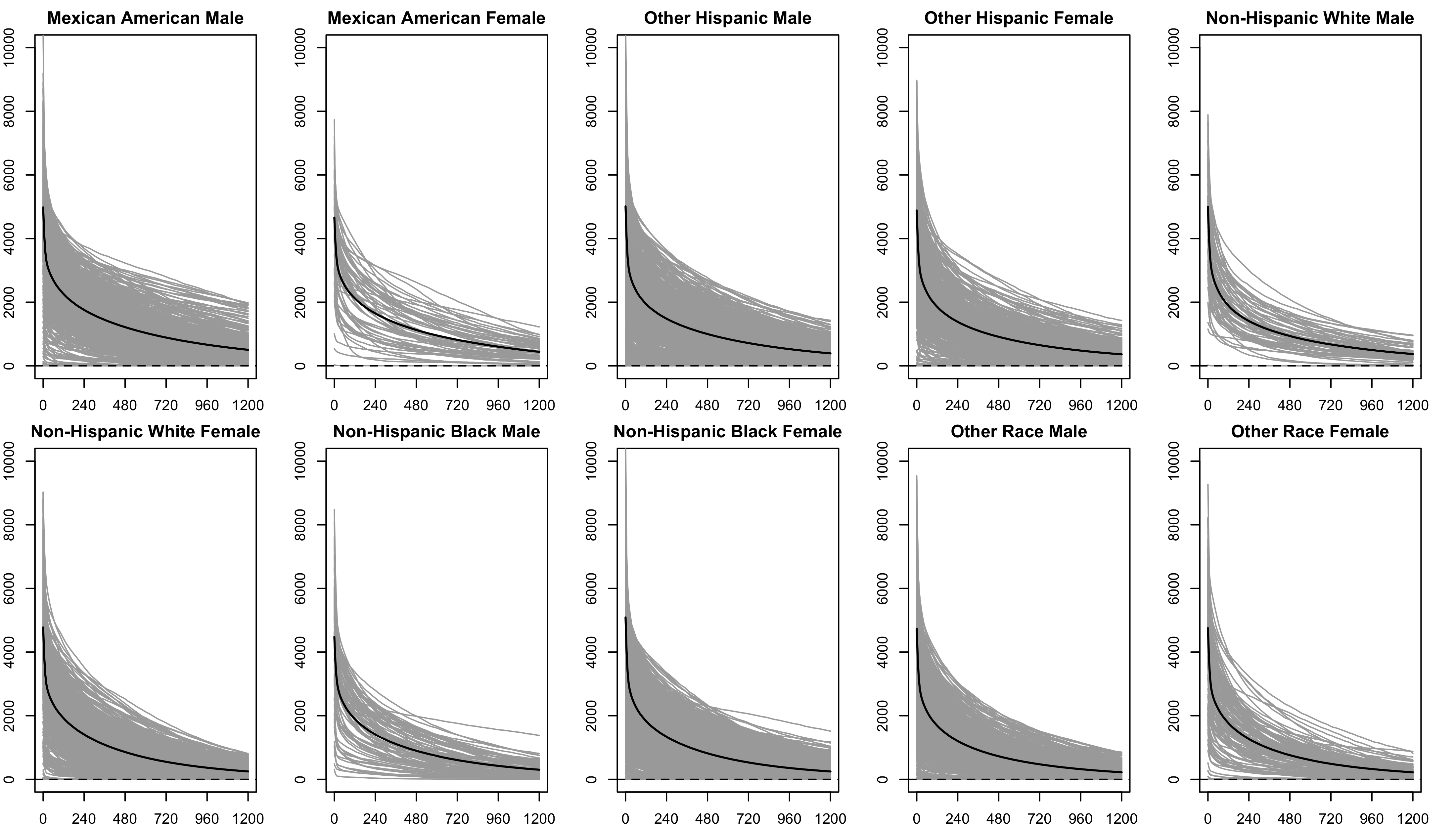}
\caption{Occupation time functions for the 10 groups of participants. The black curves are the associated average function.} 
\label{occup}
\end{figure}

We first remove the outliers based on the $\ell^2$-norm of the raw activity curves for the ten groups. The outliers are defined as those beyond the interval $[Q_1-1.5\times \mbox{IQR}, Q_3+1.5\times \mbox{IQR}]$. where $Q_1,Q_3$ are the first, third quantile of the function $\ell^2$-norm, and $\mbox{IQR}=Q_3-Q_1$. We transform the raw activity functions into occupation time functions. Occupation time refers to the duration a stochastic process spends within a particular set, such as exceeding a threshold level ``$a$". Let $\{X(t)\colon t \in [0, T]\}$ be a measurable stochastic process representing the complete observed trajectory of a raw activity curve for a given subject. The occupation time that $X(t)$ spends above the threshold $a$ is defined as: $L(a) = \text{Leb}\left(\{t \in [0, T] \colon X(t) > a\}\right)$, where $\text{Leb}$ denotes the Lebesgue measure on the real line. The index ``$a$" varies over the range of activity levels being studied. As a function of ``$a$", $L(a)$ is bounded between 0 and $T$, is monotonically decreasing, and is right-continuous by definition.  
The original occupation time functions for the ten groups are displayed in Figure \ref{occup}. Then we take the square root transformation for the occupation functions and log-transformation for chronological age to stabilize variance. The log-transformed age and the square-rooted occupation time functions are then used as response and covariate variables respectively.

\subsection{Estimation Performance and Comparison}
Here, we find that the estimation of models established for Mexican American females (target domain) is significantly improved by the data collected from Mexican American males and Non-Hispanic white males (informative source domains). To evaluate the estimation performance, we applied the Monte Carlo cross-validation procedure (see \citet{Picard01091984}). We split the entire dataset into random portions for the training set (75\%) and the test set (25\%). The target model is estimated based on the training set, which is then used to predict the test set. The steps are repeated for 1000 times. 
Here we defined the RMSE as follows:
$$\mbox{RMSE}=\left\{\frac{1}{N}\sum_{n=1}^N(y^{(b)}_n-\hat{y}^{(b)}_n)^2\right\}^{1/2},$$
where $y^{(b)}_n$ is the responses in the $b$-th bootstrap repetition, and $\hat{y}^{(b)}_n$ are their predictions. 
The box-plots of the prediction RMSE are shown in Table \ref{rmse-real}.

\begin{table}[ht]
\centering
\captionsetup{justification=centering}
\caption{Average estimation RMSEs. $\mbox{NA}$ indicating fitting with a model with a zero coefficient function. The values in the parenthesis pertain to the standard deviation of the estimation RMSEs.}
\begin{tabular}{>{\centering\arraybackslash}p{0.8in}|>{\centering\arraybackslash}p{0.8in}|>{\centering\arraybackslash}p{0.8in}|>{\centering\arraybackslash}p{0.8in}}    
\hline 
    Methods  &TS  & OR  & NA  \\
    \hline
    RMSE & 0.376(0.055) & 0.388(0.059) & 0.417(0.056)\\
      \hline 
\end{tabular}
\label{rmse-real}
\end{table}

\begin{figure}[ht]
\center
\includegraphics[width=6.2in]{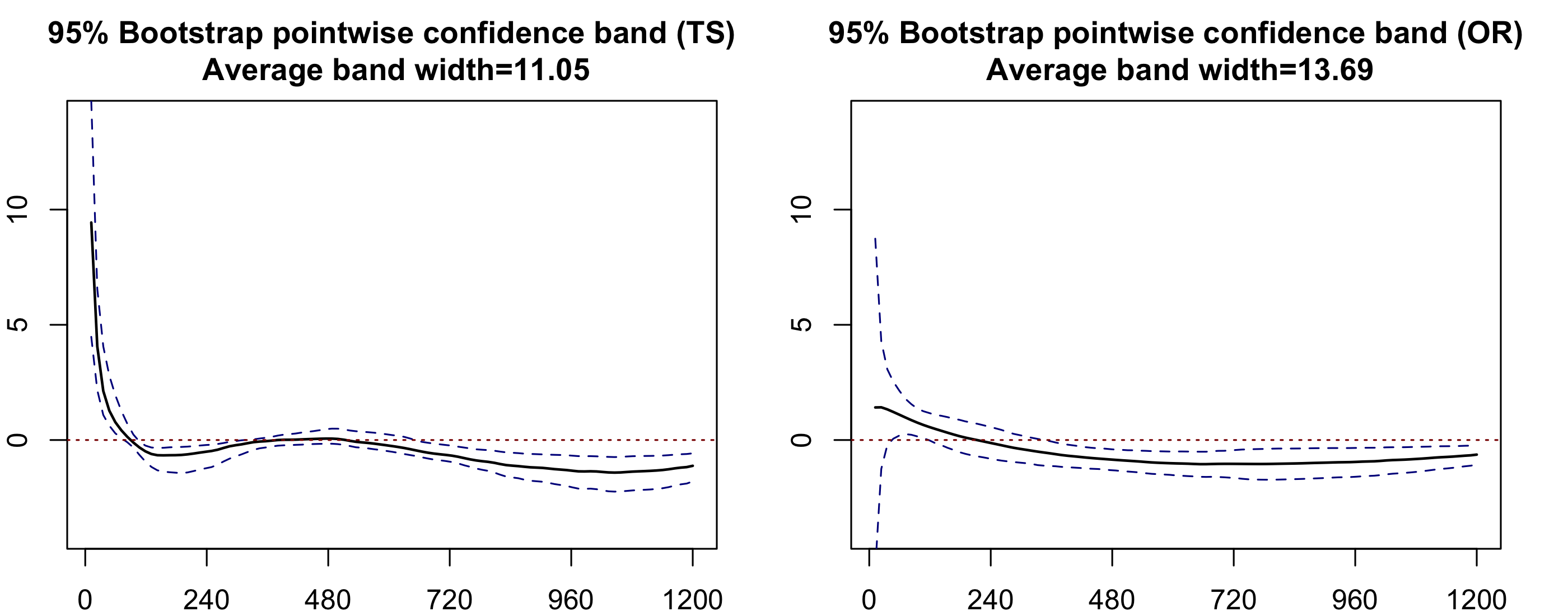}
\caption{Estimated coefficient function (black line) and the 95\% bootstrap pointwise confidence bands (dashed lines).} 
\label{cb}
\end{figure}

We also construct confidence bands to quantify the uncertainty of the estimated functional coefficient in the target domain. To achieve this, we develop a residual-based bootstrap procedure (Algorithm \ref{al3}, see e.g., \citet{gonzalez2011bootstrap}) that resamples the fitted residuals and refits the model repeatedly to capture the sampling variability of $\hat{\beta}^{(0)}(t)$. This approach allows us to construct pointwise confidence intervals across the entire domain of $t$. The resulting 95\% bootstrap confidence bands provide a visual representation of the estimation uncertainty and are displayed in Figure \ref{cb}.
\begin{algorithm}[H]
 \caption{Bootstrap Confidence Band}
  \begin{algorithmic}[1]
   \State {Fit the functional linear model and obtain residuals:
$\widehat{Y}_i = \int X_i(t) \hat{\beta}^{(0)}(t) dt$, $\hat{\epsilon}_i = Y_i - \hat{Y}_i$.}
  \State  {Resample residuals  $\hat{\epsilon}_i^*$ with replacement and generate new responses: $Y_i^* = \widehat{Y}_i + \hat{\epsilon}_i^*$.}
   \State  {Refit the model using bootstrapped responses to get  $\hat{\beta}^*(t)$.}
   \State {Repeat steps 2–3 many times (e.g., 1000 times) to obtain an empirical distribution of  $\hat{\beta}^*(t)$.}
   \State {Compute pointwise percentiles (e.g., 2.5\% and 97.5\%) to construct confidence bands.}
      %\State \Return {$\bm{b}^{(m+1)}$.}
  \end{algorithmic}
  \label{al3}
\end{algorithm}
We found that the transfer learning approach improves the effect size of the target model from 7\% to 10\% and also narrows the confidence bands of the coefficient function. The pattern of the coefficient function is significantly changed by the source domains. For example, it turns out that the period from 0 to 100 of the occupation time function has a significant effect on the response, whereas the ordinary method leads to the opposite conclusion.

\section{Conclusions}
\label{s6}
We have developed a novel transfer learning framework for functional linear models that leverages coefficient shape similarity to transfer structural information across domains. By shifting the focus from strict coefficient homogeneity to coefficient shape homogeneity, our framework substantially broadens the scope of transferable knowledge, offering a more robust and interpretable foundation for cross-domain inference. This flexibility enables the integration of heterogeneous yet structurally related sources, thereby improving estimation accuracy in the target domain.

The proposed approach proceeds in two main stages. First, we estimate a shared coefficient shape by aggregating information from both the target domain and source domains. In the second stage, we refine this estimated template by multiplying it with a scale parameter to obtain the final estimator of the target coefficient. We rigorously study the convergence properties of the resulting final estimator under both non-random and random designs. Theoretical analysis reveals that the convergence rate is influenced by the sample sizes of different domains, the spectral decay rate of covariate covariance operators, the magnitudes of coefficients, and the shape divergence between the source and target domain coefficients. To address the challenge of unknown informativeness, we develop a source domain selection procedure that identifies informative source domains without relying on prior knowledge. The practical effectiveness of our method is demonstrated through extensive simulation studies and real data analysis using NHANES physical activity data.
The proposed framework, grounded in the concept of coefficient shape homogeneity, provides a flexible foundation for broader applications and can be extended to other functional linear models, including functional autoregressive models and functional factor models. These extensions are nontrivial and will be pursued in future work.

\bibliographystyle{agsm}
\bibliography{transfer-learning}

\end{document}